\documentstyle[12pt]{article}
\begin{document}
\begin{center}
{\Large Can Quantum Wormholes really help set $\Lambda \rightarrow 0$ ?}\\

\bigskip
{\bf D.H. Coule and D. Solomons}\\
Department of Applied Mathematics\\
University of Cape Town \\
7700 Rondebosch, RSA.
\bigskip
\begin{abstract}
We find the quantum analogues of wormholes obtained by
 Carlini and Miji\'{c} (CM), who
analytically continued closed universe models. The CM
requirement that the strong energy condition
($\gamma >2/3$) be satisfied is
shown to be consistent with the Hawking-Page conjecture for
quantum wormholes as solutions of
the Wheeler-DeWitt equation. The presence of a cosmological constant
$\Lambda$ violates such a condition and so prevents wormholes
occurring. It is therefore inconsistent to invoke these wormholes to
make $\Lambda$ a dynamical variable: used in arguments which
suggest $\Lambda \rightarrow 0$.\\
We analyse instead a simple model with just $\Lambda$ present.
 Differing  results are
obtained depending on the boundary conditions applied. In the
Euclidean regime only the Hartle-Hawking boundary condition
gives the factor $exp(1/\Lambda)$ but is badly behaved for
negative $\Lambda$ . Tunneling boundary conditions
suggest an initially large value for $\Lambda$. Whereas in the
Lorentzian region all boundary conditions suggest an initially
large value of $\Lambda$ for spatial
curvature $k=1$. This differs from the previously
obtained result of Strominger[28] for such models.

\end{abstract}
\end{center}
\newpage
\section{Introduction}
One possible solution to
the cosmological constant $\Lambda$ problem
that  has attracted a lot of interest is due
to the idea that wormhole solutions can lead $\Lambda$ to become a
dynamical variable with a distribution function $P(\Lambda)$ [1]
- for a review of this proposal see eg. ref.[2]. It is
suggested that this
function is peaked, due to De Sitter instantons, with  the Baum-Hawking
factor $P(\Lambda)\sim exp(1/\Lambda)$ [3,4] so predicting $\Lambda
\rightarrow 0$ [1]. Wormholes are used in two distinct ways in such arguments,
firstly they are used to justify why $\Lambda$ should be treated as a
dynamical quantum variable instead of a usual classical variable. This
is the most important aspect, as it allows one to make
 predictions  of the possible values of $\Lambda$.
 Wormholes have further
been used in connecting many universes together which produces a further
exponentiation [1] i.e.
\begin{equation}
P(\Lambda) \sim exp\left (exp \left ( 1/\Lambda \right )
\right )
\end{equation}
This is only useful if the first factor $\sim 1/\Lambda$ is correct. In
other words, this aspect of wormholes only exaggerates any underlying
behaviour.

Carlini and Miji\'{c} [5]  have greatly expanded the number of possible
wormhole solutions by considering an analytic continuation of
closed Friedman-Robertson-Walker (FRW) universes. For a perfect fluid
equation of state, closed universes require that the strong
energy condition be satisfied i.e. $\gamma>2/3$. By using an
adaption of an approach used by  Ellis and
Madsen [6] the value of $\gamma$ can be fixed. This in turn can
be formulated in terms of a scalar field model
 with a rather complicated potential. This enables you to see
the significance of the strong energy condition in setting up
closed universe models.  It is the
continuation of this theory to the Euclidean domain which then gives
wormhole solutions. One drawback
of this approach is that the analytic continuation is rather
ad hoc with the matter and gravitational parts being
 treated differently [5].

 In a different vein,
because the number of known wormhole solutions had appeared so
limited, Hawking and Page (HP) considered that  solutions
of the Wheeler-DeWitt (WDW) equation could more generally
 represent wormholes [7].
 For such
wormholes they suggested that the quantum mechanical wavefunction $\Psi$
 decay exponentially for large scale
factor $a$ so as to represent Euclidean space, and that
$\Psi$ be well behaved as  $a\rightarrow 0$: so that
no singularities are present.

By starting with a scalar field with a potential that fixes
a value of $\gamma$  we intend deriving the corresponding WDW equation.
Because the WDW equation is independent
of the Lapse chosen, the Euclidean regime is already included in the formalism.
There is thus no need in making any arbitrary continuation and the quantum
versions of the Carlini-Miji\'{c}
 wormholes can be found. Such solutions
are found to obey the Hawking-Page behaviour when the matter obeys
the strong energy condition $(\gamma >2/3)$. This is the same
condition that CM required for a closed universe; that could then be
analytically continued to a wormhole [5].

Another point raised by the CM wormholes is whether spatial
curvature $k=1$ is really necessary for a wormhole. Although all
previously considered wormholes have this feature it is not strictly
necessary for a closed universe. Certain unstable fields are known
to cause recollapse [8,9] but for simplicity the
case of a -ve cosmological constant can be considered.

We first review the classical closed universe models from which we are
going to derive their corresponding WDW equations. First we take
 a bulk matter source with a perfect fluid equation of state $p=(\gamma
-1)\rho$, ($p$ and $\rho$ are the pressure and energy density respectively).\\
Working in a Lorentzian metric
\begin{equation}
ds^2=-a^{4-3\gamma}dt^2+a^2(t)d\Omega^2_3
\end{equation}
where we are using the same lapse to aid calculations
as in Refs.[5], the scale factor $a$ is given by
\begin{equation}
a(t)=\left[ a_o^{3\gamma-2}-(1-3\gamma/2)^2t^2\right ]^{1/(3\gamma-2)}
\end{equation}
With $a_o$ an arbitrary constant which is the maximum size of the
Robertson-Walker universe.
 Using the same
approach as in Ellis-Madsen [6]
we can convert to a scalar field $\phi$ whose trajectory is given by
\begin{equation}
\phi=\frac{2}{3\gamma-2} \gamma^{1/2}tanh^{-1}\left [\frac{3\gamma-2}
{2a_o^{(3\gamma-2)/2 }}t\right ]
\end{equation}
The solution have a potential of the form
\begin{equation}
V(\phi)=\frac{2-\gamma}{2a_o^2}cosh ^{\frac{6\gamma}{(3\gamma-2)}}
\left[ \frac{(3\gamma-2)\phi}{2\sqrt{\gamma}}\right ]
\end{equation}
Notice the difference in the preceding expressions with their
Euclidean counterparts in refs.[5]. These expressions are valid for
a closed universe analogous expressions could be obtained for $k=-1$.
 By using eq.(3) and (4)
 the potential can be rewritten as a function of the
scale factor.
\begin{equation}
V(\phi)\equiv V(a)=\frac{V_m}{a^{3\gamma}}
\end{equation}
where the constant $V_m=(1-\gamma/2)a_o^{3\gamma-2}$.

\section{Quantum Wormholes}

We consider solutions of the Wheeler-DeWitt equation, see e.g. [10-12]
\begin{equation}
\left ( \frac{\partial}{\partial a^2}+\frac{p}{a}\frac{\partial}{
\partial a}-\frac{1}{a^2}\frac{\partial}{\partial \phi ^2}-ka^2
+a^4V(\phi)
\right ) \Psi(a,\phi)=0
\end{equation}
$p$ is a factor ordering correction and $k$ the spatial curvature $\pm1,
0$.\\
By using the potential $V(\phi)$ which corresponds to a
fixed value of $\gamma$  allows the WDW equation to be separable.

The WDW equation simplifies to:
\begin{equation}
\left ( a^2\frac{d^2}{da^2}+pa\frac{d}{da}+q^2+V_ma^{6-3\gamma}
-ka^4 \right ) \Psi(a)=0
\end{equation}
\begin{equation}
\left (\frac{d^2}{d\phi ^2}+q^2\right )\Psi(\phi)=0
\end{equation}
with $q$ the separation constant. \\
We can get some idea as to when a Euclidean domain occurs at large $a$
by considering the sign of the potential $U$ in the analogous equation [10]
\begin{equation}
\left ( \frac{d^2}{da^2} +U\right ) \Psi(a)=0
\end{equation}
When $U>0$ oscillating solutions occur which represent Lorentzian metrics.
So in order to obtain a wormhole (an asymptotically Euclidean
regime for large scale factor) we require $U<0$ .
Returning to our  eq.(8), and setting the
unimportant in this regard factor $p=0$, this occurs for
\begin{equation}
V_m a^{4-3\gamma}-ka^2<0
\end{equation}
  Therefore for the usual case of
positive potential($V_m > 0$) we require
$2>4-3\gamma$ i.e. $\gamma >2/3$ and $k=1$
for such behaviour .
If we have a negative potential ($V_m< 0$) then this is reversed
; $\gamma<2/3$ and  $k=\pm 1$ can  give such a solution. This is an example
that shows that $k=1$ is not strictly necessary to obtain wormhole solutions
 - at least in the CM and HP sense.\\
 We have ignored  the term involving the
separation constant $q/a^2$ which  would only be important at small $a$.
This strong energy condition is the same as that obtained by CM for the
occurrence of wormhole solutions.
 Recently  Kim and Page [13]  seem to find also  that quantum
wormholes are incompatible with a
 cosmological constant. However we  find here   a  stronger condition
: matter sources violating the strong energy condition are
incompatible with wormholes obeying the HP conditions.
 The presence of any matter source
with $\gamma<2/3$ will eventually dominate for large $a$ and prevent
the Euclidean wormhole .\\
 A quantum wormhole also requires suitable
behaviour for small $a$. As $a\rightarrow 0$ we can ignore the $ka^4$
term since $4>6-3\gamma$ when $\gamma>2/3$. In this case the WDW equation
simplifies to a Bessel equation with solution
\begin{equation}
\Psi(a)\sim J_{iq}(a^{3-3\gamma/2}\sqrt {V_m}) + Y_{iq}(a^{3-3\gamma/2}
\sqrt {V_m})
\end{equation}

Using the asymptote $J_{\nu}(z) \sim z^{\nu}$ as $z\rightarrow 0$
enables the solution to be expressed as
\begin{equation}
\Psi(a)\sim exp\left \{iq(3-3\gamma/2)lna\right \}
\end{equation}
The other Bessel function would simply have a (-) sign in the exponent
of eq.(13) since $Y_{\nu}(z)\sim -z^{-\nu}$. The foregoing argument would
proceed in the same fashion.
We have also taken $p=1$ although the analysis could easily be extended
to include other cases.\\
The problem now is that as $a\rightarrow 0$ the $ln a \rightarrow -\infty$
causes infinite oscillations to occur, the
wavefunction  cannot be regarded as a wormhole in its present form as the
oscillations represent a singularity [7]. The full solution is:
\begin{equation}
\Psi(a,\phi) \sim exp\left \{iq\left [(3-3\gamma/2)ln a +\phi \right ]
\right \}
\end{equation}
By integrating over the separation constant we can eliminate this
singularity at the origin. This is the same as performing a Fourier
Transform [14,15] or  like forming a wave-packet solution [16].
Now the integral
\begin{equation}
\Psi\sim \int exp\left \{iq\left [ (3-3\gamma/2) ln a +\phi\right ]
\right \} dq
\end{equation}
Is of the form $\int exp(ixt)dt$ which by means of the Riemann-
Lebesgue Lemma (see eg. ref.[17])
 $\rightarrow 0$ as $x\rightarrow \infty$. The wavefunction
is now, in effect damped as $a\rightarrow 0$ or $\phi \rightarrow \infty$.
Such solutions now
obey the HP regularity condition and  can
be taken to be wormholes.

We now return to the full WDW equation;
although it  has been simplified to an ordinary differential
equation it is still not
straightforward to obtain analytic solutions for all
$\gamma$. You could proceed by finding
approximate WKB solutions;  but instead
we consider only certain values of $\gamma$ that are exactly
soluble -  this still
enables us to emphasize  important properties
 that any solution in the range $2\geq \gamma >2/3$ will have.
We should point out that throughout this section we are only interested
in the existence of wormhole solutions, and set arbitrary coefficients
accordingly. In theory such coefficients should be determined by the
 boundary conditions applied. The usually applied boundary conditions
of Hartle-Hawking [18] or  the Tunneling one [19], would in general contradict
such solutions - they also generally require $q=0$ - see ref.[20]. We
have already seen that the wave-packet type solution is more in keeping
with  the HP conditions for a quantum wormhole.

a)$\gamma=2$ or $V(\phi)=0.$\\
 This is the minimally coupled
massless scalar field. The solution of equations (8) and (9) is cf.ref.[7]
\begin{equation}
\Psi \sim e^{iq\phi} \left \{ J_{iq/2}(ia^2/2)+ Y_{iq/2}(ia^2/2)\right \}
\end{equation}
This has an oscillation phase for $a<q$ and an exponential fall off
for $a>q$ : cf. Fig.(1). Let us
call this type(I) behaviour of the wavefunction.
 Classically this is a closed universe with a forbidden
region when $a>q$, can this type(I) wavefunction
behaviour be considered a wormhole as claimed by
Hawking and Page [7]?.  A classical
wormhole is known to occur for a totally imaginary scalar field. This
corresponds to a change of sign in the $\partial^2/\partial \phi^2$ term
. In order to keep plane wave solutions in $\phi$ the separation constant
would change sign $q^2\rightarrow -q^2$ and the solution is
\begin{equation}
\Psi \sim e^{iq\phi}\left \{ K_q(a^2/2)+I_q(a^2/2)\right \}
\end{equation}
This no longer has an oscillating domain and is instead Euclidean for
all $a$ cf. Fig.(2). Although if the Modified Bessel  function $K$ is chosen
the
wave function is divergent for $a\rightarrow 0$. We can call such
a wavefunction type(II) behaviour and is the sort of wavefunction that
corresponds to situations that classically would allow wormholes.

b) $\gamma=4/3$: Radiation or a conformally coupled scalar field.\\
 A second example is that of radiation $\gamma=4/3$ which allows eq.(8)
to be written as cf.ref.[21]
\begin{equation}
\left (\frac{d^2}{da^2}+V_m-a^2 \right )\Psi(a)=0
\end{equation}
This is in the form of a Parabolic cylinder equation with solutions in
terms of confluent hypergeometric functions -see eg.[22]
\begin{eqnarray}
\Psi(a)& \simeq & exp(-a^2/2)\;_1F_1\left ((1/4(1-V_m);1/2;a^2\right )
\nonumber \\
& + &  exp(-a^2/2)\;_1F_1\left ( 1/4(3-V_m);3/2;a^2 \right )
\end{eqnarray}
For $V_m>1$ we  get an oscillation for small $a$ with the Euclidean
regime again for large $a$ see Fig.(1)- this is again a type(I)
wavefunction. As  $V_m$ is
lowered  the Euclidean
regime gets closer to the origin. With $V_m\leq 0$ it is totally Euclidean
and becomes type (II) see Fig.(2). This is now a negative
 energy density which should classically
be able to support a wormhole solution.\\

Whether type(I) or the more restrictive type (II) behaviour
 is the correct description for
the wavefunction to
describe wormholes there is something of a problem here. We  would expect
a Euclidean regime to occur at small size and not to correspond to a
region beyond the size of a Lorentzian one. The type(II) description has
a Euclidean region at small size, but the presence of any matter
 source violating the strong energy
condition will make it  have Lorentzian behaviour for larger $a$. It would
 no longer  obey
the HP description and  requires matter sources which anyway
have classical wormholes solutions. If the type (I)  wavefunction correctly
describes a
 wormholes then the hope of Hawking and Page, that arbitrary
matter sources could have wormhole solutions  would appear not be be true.\\
Can we not simply include a $\Lambda$ term together with a matter source with
$\gamma >2/3$ ? After all the $\Lambda$ term only dominates at large size
and many classical wormholes can be constructed also with $\Lambda$ present:
this results in the wormhole being attached to Euclidean De Sitter space, see
e.g. [23,24]. In the WDW equation however, the De Sitter regime is necessarily
Lorentzian at large size.  The wormhole behaviour at small
size is also Lorentzian and the presence of $\Lambda$ can only potentially
result in a barrier between these two regions. The solution would instead
represent a tunnelling event between these two regions.  In order to get the
important $\exp (1/\Lambda)$ factors we need rather to attach the wormhole to
Euclidean De Sitter spaces. In this sense the solutions of the
WDW equation are less general
than required in Coleman's approach [1]
since they cannot incorporate such geometries.
\footnote {There is an interesting analogy with the problem found
by Verbin and Davidson [24]
 in the case of the conformally coupled wormhole. When a potential
$\lambda \phi^4$ was introduced the wormhole cannot exist
(its size $<<$ Planck size)  if $\Lambda
\neq 0$.}

The main conclusion of this section
is that quantum wormholes are prevented by the presence of matter
sources violating $\gamma <2/3$ (which includes a $\Lambda$ term).
   With this contradiction it is therefore uncertain what role such
solutions of the WDW equation
 can have in the setting to zero of the cosmological constant.
One approach out of this impasse , the 3rd quantization, is
to simply piece together many separate wavefunctions each representing a
separate universe [25].  Instead we stick with the usual (2nd quantized)
WDW equation and
next consider the case, without wormholes,
of just a cosmological constant $\Lambda$ present.\\

\section {Cosmological constant model}
 When only a  cosmological constant is present the WDW equation
 takes the simplified form.
\begin{equation}
\left ( \frac{d}{da^2}-U \right )\Psi(a)=0
\end{equation}
Where
the WDW potential $U$ for  a closed $k=1$ universe is given by
\begin{equation}
U=a^2-\Lambda a^4
\end{equation}
The potential is sketched in Fig.(2). This has been studied by
many authors especially as the case of quantum tunnelling to a
Lorentzian universe [26]. We follow particularly the analysis of
Lavrelashrili  et.al. [27] and Strominger[28].
The WKB solutions have the form cf.[27]
\begin{equation}
\Psi=\frac{1}{\sqrt{|U(a)|}}exp\left (\pm\int \sqrt{U(a)}da\right )
\end{equation}
where the `action' $S=-\int\sqrt{U}da$ is given by
\begin{equation}
-\int a(1-\Lambda a^2 )^{1/2}da=\frac{(1-\Lambda a^2)^{3/2}}{3\Lambda}
\end{equation}
Taking the limits between $a=\Lambda^{-1/2}$
and  $a=0$  gives the solutions
\begin{equation}
\Psi_{\pm}\sim exp(\pm 1/\Lambda).
\end{equation}
The (+) sign corresponds to the Hartle-Hawking (HH)[18]
 boundary condition $exp(-S)$
and the (-) sign to the tunnelling one $exp(-|S|)$- see
for example [11,12].\\
 If we assume that the probability
of having a specific $\Lambda$ is $P(\Lambda)\sim \Psi^2 \sim exp (\pm 2
/\Lambda )$
 then the two approaches predict $\Lambda \rightarrow 0$ and
$\Lambda \rightarrow \infty$ respectively.
For the HH case we appear to get the suppression of $\Lambda$ but the
opposite for the tunnelling case. However the tunnelling occurs through
the barrier to $U=0$ where $a^2\sim 1/\Lambda$. When the barrier is
small i.e.  when $\Lambda $ is
large tunnelling is enhanced: this is somewhat
analogous to the application of an electric potential to an atom which
allows electron  to tunnel away. In this case a large value of $\Lambda$
is enhancing the possibility of the universe tunnelling into existence.
\\
According to Strominger [28] because the scale factor $a$ today is very large
the value of $\Lambda\sim a^{-2}$ is very small as required to fit
observation. This does not however agree with the usual interpretation
of quantum cosmology which is that of predicting initial conditions. As
quantum  tunneling is expected to occur to an initial size of roughly
Planck dimensions the initial value of $\Lambda$ is correspondingly
big $\sim 1$ in Planck units. If instead the initial scale factor
was large (and so
the initial value of $\Lambda
\sim$ small) it would mean that the Euclidean domain would extend to
large sizes. It would  then be inconsistent with the present structure
of space-time which appears Lorentzian down to at least sizes of $\sim
10^{-20}$ meters.\\
We consider next a problem that has arisen for the case of a
$-ve$ cosmological constant. The equivalent expression to
eq. (23) is
\begin{equation}
S= -\frac{(1+|\Lambda|a^2)^{3/2}}{3|\Lambda |}
\end{equation}
 It is no longer clear
what integration  limits have to be placed on $a$. Choosing them from
$a$ to $a=0$ gives the solutions

\begin{equation}
\Psi\sim exp\pm \left[\frac{(1+|\Lambda | a^2)^{3/2}}{3|\Lambda |}
-\frac{1}{3|\Lambda|}\right ]
\end{equation}
If we keep track of the signs then the (+) one corresponds to the HH case
and will be dominated by large
\begin{equation}
\sim \left [ \frac{(1+|\Lambda|a^2)^{3/2}}{
3|\Lambda |}-\frac{1}{3\Lambda }\right ]
\end{equation}
 i.e. by $\sqrt{|\Lambda |}a^3$ large.\footnote{
If we had not subtracted the part corresponding to $a=0$ we would
also find a divergence when $|\Lambda|\rightarrow 0$}.
 This is the problem
found by  Lavrelashrili et. al.[27]
\footnote{It is not necessary, as done in ref.[27] to include matter fields
or to consider a third quantized theory to obtain this
dominant $\sqrt{|\Lambda|}a^3$ factor.}
 It is however uncertain that this makes any sense and
is rather an artifact of the HH wavefunction been peaked around the
exponentially increasing solution. cf. Fig.(2) in Ref.[26].\\
There is another reason to discount this solution. If the cosmological
constant was absent the wave function would be\\
\begin{equation}
\Psi \sim exp \left ( \pm a^2/2 \right )
\end{equation}
If we choose the ($+$)  sign there is  a contradiction with
our notions of classical behaviour since the universe would apparently
prefer to have large size. Rather the other sign is more correctly
peaked around $a=0$.\\
 The other (-) sign  solution in eq.(26)  formally appears to
predict $\Lambda \rightarrow 0$ if $a\neq 0$. But since there
 is no barrier to tunnel through the tunneling  condition will
simply imply that the universe stays at the origin $a=0$ and $ \Lambda$
is left undefined. One can seemingly obtain large or small $|\Lambda|$
depending on how one applies the boundary condition (the limits of
integration in eq.(22) ). When considering a -ve $\Lambda$ it
seems that we should conclude that the universe will
wish to stay at the origin and no predictions about $\Lambda$ should
be drawn from the factor $exp{\sqrt{|\Lambda|}a^3}$. We limit our
discussion to the case of positive values of $\Lambda$ again from now.\\
In this Euclidean region  we have found that the possible values of $\Lambda$
depend upon the choice of boundary conditions.
This point has also been mentioned  by Kiefer [29] in the context of wave
packet solutions to the WDW equation.
 Because the choice of boundary conditions
is not known a priori, it seems that to simply choose the boundary condition
that solves the cosmological constant problem is merely to pass  the
problem down the line.
 What is required is a measure of solutions to the WDW equation
which give either large or small $\Lambda$. Any possible wave function
\begin{equation}
\Psi \sim \alpha exp(1/\Lambda)+\beta exp(-1/\Lambda)
\end{equation}
 ($\alpha, \beta $ arbitrary complex coefficients)
 will have the critical behaviour
at $\Lambda=0$ for $\alpha$ small.  However, it
would make $D>>1$  (defined in ref.[30]) but
according to the measure given in ref.[30]  a typical wavefunction
has $D<1$: it  is therefore more like  a tunneling one i.e. of the form
$exp(-|S|)$ .
If this measure is correct
it would suggest that  a large value of $\Lambda$ should be
expected.
Until a choice of boundary conditions can be justified,  or the
 measure of  solutions is known,  we cannot know if such
Baum-Hawking  factors
are important for determining  the likely cosmological constant.  \\

 We consider next what happens when the universe starts
in a Lorentzian region where the WKB wave functions have the oscillating
behaviour $\Psi\sim exp(\pm iS)$.
The exponents in the terms  $exp(\pm iS)$ no longer have any critical
influence, but instead
the pre-factor contains any dominant behaviour.
We do however have to exclude the Euclidean regime from the
expression for the action cf. eq.(23) i.e. the lower
limit in the integral is taken to be $a=\Lambda^{-1/2}$. Otherwise
we would simply introduce the factors $exp(\pm 1/\Lambda)$ again
and reach similar conclusions.\footnote {
 For the
same reason these factors
appear when you wish to normalize the wavefunction as $a\rightarrow 0$
-see e.g.[31].}
\\ Typically the wavefunction has the form
\begin{equation}
\Psi \sim \frac{1}{a\sqrt{a^2\Lambda-1}}\left (e^{iS}+e^{-iS}\right )
\end{equation}
There is a similar peak around $a^2\Lambda\sim 1$ as the WDW potential is
zero. For $a$ fixed and $a^2\Lambda>>1$  then $\Psi^2\sim 1/\Lambda$ so
that larger values of $\Lambda$ are suppressed inversely.
 These are again
initial conditions to be followed by classical evolution,
and it would appear correct to assume the
quantum behavior made predictions for an initially small universe. The
initial value of $\Lambda $ would therefore appear large which would
produce an inflationary phase. This prediction occurs for both HH
and tunneling boundary conditions since they only determine
which combination of $exp(iS)$ and $exp(-iS)$ to take.
There is a heuristic reason to see this: in the Lorentzian
region the HH and tunneling solutions  look almost alike
( damped oscillations) and so should not differ much in their predictions.
Contrast this with their behaviour in the Euclidean regime - see eg.
Fig.(11.2) in ref. [12].\\ \footnote{ We  should
perhaps be careful and say the
prediction is not strongly dependent on boundary
conditions since there might be more unusual ways
of imposing them cf. Ref.[32]. Note the slight discrepancy with
Cline[32] who using  a Lorentzian path integral
approach concluded that $P(\Lambda)\sim 1$ so that any
$\Lambda$ is equally likely. In the Euclidean
region he found that
only special boundary conditions gave the
$exp(1/\Lambda)$ factor - agreeing with us.}
Similar predictions could be made if we considered a spatially
open $k=-1$ model together with a -ve $\Lambda$. This has Lorentzian
behaviour for small $a$ beyond which is a Euclidean regime. The peak
would be around $a^2|\Lambda |\sim -k$. Notice  how the
spatial curvature is crucial for any predictions about $\Lambda$ . If we
set $k=0$ then $P(\Lambda) \sim 1/(a^4\Lambda)$ and we would obtain the
prediction that $\Lambda \rightarrow 0$, although without
the exponential peak.

 It might appear that this property $a^2\Lambda \sim 1$
 is an artifact of using WKB solutions which are
simply blowing up at the  turning point  $U\equiv a^2-\Lambda a^4=0$.
However exact solutions of the WDW equation can be found and this behaviour
remains. For example the equation [33]
\begin{equation}
\left (\frac{d^2}{da^2}+\frac{p}{a}\frac{d}{da}-U\right )\Psi(a)=0
\end{equation}
has solutions
\begin{equation}
\Psi\sim a^{\frac{1-p}{2}}\left \{J_{(1-p)/4}(\sqrt{-U}a)+
Y_{(1-p)/4}(\sqrt {-U}a)\right \}
\end{equation}
Since both $J_{\nu}(x)$ and $Y_{\nu}(x)$
 both decrease for increasing $x$ they
both take their maximum value when $x=0$ and as we require $a>0$ this
occurs for $U=0$ , so again when $\Lambda \sim 1/a^2$. For $p\neq 1$ this
is slightly modified $\sqrt {-Ua}\sim small$.\\
The addition of additional matter fields is likely to round off this
spike at $\Lambda \sim 1/a^2$ cf. ref.[34].
We see that the initial value of $\Lambda$ is expected to be large in
the Lorentzian regime provided $a$ is not large in Planck units. If the
initial size of the universe is `big' $\sim 1\;cm$ the
probable value of $\Lambda$ is
smaller but still huge compared to its present value cf. Ref.[28].\\
Let us finally try to understand the Euclidean regime results
(when $U\geq0$)
in terms of the
solutions  eq.(31). For factor ordering $p=1$ the solutions simplify
to
\begin{equation}
\Psi\sim K_0(aU^{1/2})+I_0(aU^{1/2})
\end{equation}
Since $K_0(x)\rightarrow \infty $ as $x\rightarrow 0$ it picks out
the $U=0$ or $\Lambda\sim 1/a^2$ case. The other Bessel function
$I$ increases with increasing $aU^{1/2}\equiv a^2(1-\Lambda a^2)
^{1/2}$ so is maximized for $\Lambda=0$,
or  negative $\Lambda$ if we allow it. This is as expected since the
tunneling boundary condition is the decaying solution $K$ and the
HH one a mixture of both $I$ and $K$- see eg.ref.[35].
\section{Conclusions}
Using the Carlini-Miji\'{c} approach of fixing $\gamma$ we have found
that quantum wormholes occur only for matter sources that
obey the strong energy condition ( the opposite to
what is required for inflationary behaviour). The solutions obtained
also obey the HP description for $\Psi$ to describe a wormhole: they
are asymptotically Euclidean and well behaved (if we take a Fourier
transform) as $a\rightarrow 0$. It might be argued that this has little
to do with actual matter sources.  For example
a massive scalar field $V(\phi)=m^2\phi^2$
 has a range $0\leq \gamma \leq 2$ and
so does not necessarily violate
the condition $\gamma>2/3$. This is why a quantum
wormhole could still be obtained
for the massive scalar field [7]. However if wormholes only
occur for matter sources which obey the strong energy condition it
seems a contradiction to invoke them to a situation which
violates such conditions i.e. a $\Lambda$ term.\\
This requirement that $\gamma>2/3$ is easier to satisfy than
the corresponding condition for classical wormholes
 ( that the Ricci tensor have negative eigenvalues [36]). But
 perhaps a still   more general form of quantum wormhole
is required which  might exist even
 when the strong energy condition is violated. Otherwise the (3rd quantized)
 many
universe approach, which is even more  speculative  could
be considered.\\
 Even the present HP definition of a
quantum wormhole  seems  suspect since they
can appear `more quantum' Euclidean  on large scales.
 We would  instead expect Euclidean regimes only to occur on
small  $\sim$ Planck  scales when quantum gravity is expected to
be important.\\
Instead of considering wormhole solutions of the WDW equation we
took a simple model with only a cosmological constant present.
In the Euclidean regime the initial value of $\Lambda$ is expected
large if tunneling boundary conditions or tunneling like behaviour
is correct. This would not be suitable for setting $\Lambda$ small
but would allow for an Inflation regime to proceed. If
we consider HH boundary conditions then
you can get the factor $exp(1/\Lambda)$ (or if $\Lambda$ is -ve the
anomalous
$exp{\sqrt{|\Lambda |}a^3}$ factor).
 The two behaviours are complimentary and the occurrence of one
 would prevent the other: we could not have Inflation together with
$\Lambda \rightarrow 0$ unless some other dynamical mechanism could
give the two mechanisms differing time scales cf. Ref.[37] .\\

We then considered  the purely  Lorentzian regime and found that the
predictions in this case are
 not dependent on the boundary
conditions.
\footnote{In this regard we essentially agree with the analysis of Strominger
[28], except for his conclusion that this implies $\Lambda \rightarrow
\simeq 0$.}
 We found an initial value of $\Lambda \sim k/a^2$
 and  since we expect the universe to start small, due to
quantum gravity processes, the corresponding value
of $\Lambda$ is large.
We are still left with the problem of why $\Lambda$ should be a
dynamical variable with a distribution function.\\  If wormholes
are not allowed when $\Lambda$ is present, it would be a contradiction
to then invoke them to make $\Lambda$ a dynamical variable. The 3-form
(axion) field might still work in this regard even if
its wormhole solution cannot be invoked cf.ref.[2].
\\
{\bf Acknowledgement}\\
We should like to thank Prof. G.F.R. Ellis for his continued interest
and encouragement during the course of this work.
\section{References}
\begin{enumerate}
\item S. Coleman, Nucl. Phys. B 310 (1988) p.643.
\item S. Weinberg, Rev. of Mod. Phys. 61 (1989) p.1.
\item E. Baum, Phys. Lett. 133 B (1983) p.185.
\item S.W. Hawking, Phys. Lett. 134 B (1983) p.403.
\item A. Carlini and M. Miji\'{c}, ``Spacetime wormholes as analytic
continuation of closed expanding universes'' SISSA preprint 91A (1991).
\\
A. Carlini, preprint SISSA 65/92/A (1992).
\\
A. Carlini and M. Martellini, Class. Quan. Grav. 9 (1992) p.629.
\item G.F.R. Ellis and M.S. Madsen, Class. Quan. Grav. 8 (1991) p.667.
\item S.W. Hawking and D.N. Page, Phys. Rev. D 42 (1990) p.2655.
\item L.H. Ford, Phys. Lett. 110 A (1985) p.21.
\item A. Vilenkin, Phys. Rev. D 33 (1986) p.3560.

\item J.J. Halliwell, in {\em Quantum Cosmology and Baby Universes}
eds. S. Coleman et al. (World Scientific, Singapore) 1991.

\item A.D. Linde, {\em Particle  Physics and Inflationary Cosmology}
(Harwood, Switzerland 1990).
\item E.W. Kolb and M.S. Turner, {\em The Early Universe}, (Addison-
Wesley, USA 1990).
\item S.P Kim and D. N. Page, Phys. Rev. D 45 (1992) p.R3296.
\item A. Zhuk, Phys. Rev. D 45 (1992) p.1192.
\item L.J. Garay, Phys. Rev. D 44 (1991) p.1059.
\item C. Kiefer, Phys. Rev. D 38 (1988) p.1761.
\item C.M. Bender and S.A. Orszag, {\em Advanced Mathematical Methods for
Scientists and Engineers}, (McGraw-Hill, Singapore 1984).
\item J.B. Hartle and S.W. Hawking, Phys. Rev. D 28 (1983) p.2960.
\item A. Vilenkin, Phys. Rev. D 30 (1984) p.549\\
A.D. Linde, Sov. Phys. JEPT 60 (1984) p.211\\
V.A. Rubakov, Phys. Lett. 148 B (1984) p.280.\\
Y.B. Zeldovich and A.A. Starobinski, Sov. Astron. Lett. (1984) p.135.
\item M.B. Miji\'{c}, M.S. Morris and W. Suen, Phys. Rev. D 39 (1989) p.1496
\item P.F. Gonzalez-Diaz, Mod. Phys. Lett. A 5 (1990) p.1307.
\item E. Merzbacher, {\em Quantum Mechanics} (Wiley, New York) 1961.
\item J.J. Halliwell and R. Laflamme, Class. Quan. Grav. 6 (1989) p.1839.
\item Y. Verbin and A. Davidson, Nucl. Phys. B 339 (1990) p.545.
\item V.A. Rubakov, Phys. Lett. 214 B (1988) p.503.\\
A. Hosoya and M. Morikawa, Phys. Rev. D 39 (1989) p.1123.
\item A. Vilenkin, Phys. Rev. D 37 (1988) p.888.
\item G. Lavrelashrili, V.A. Rubakov and P.G. Tinyakov, in {\em
Gravitation and Quantum Cosmology} eds. A. Zichichi et al. (Plenum
Press, New York, 1991) p.87.
\item A. Strominger, Nucl. Phys. B 319 (1989) p.722.
\item C. Kiefer, Nucl. Phys. B 341 (1990) p. 273.
\item G.W. Gibbons and L.P. Grishchuk, Nucl. Phys. B 313 (1989) p.736.\\
 L.P. Grishchuk and Y.V. Sidorov, Sov. Phys. JEPT 67 (1988) p.1533.
\item E. Fahri, Phys. Lett. B 219 (1989) p.403.
\item J. M. Cline, Phys. Lett. 224 B (1989) p.53.
\item D.H. Coule, Class. Quan. Grav. 9 (1992) p.2353.
\item T. Banks, Nucl. Phys. B 309 (1988) p.493.
\item D.N. Page, in {\em Proceedings of Banff summer research institute
on Gravitation} eds. R. Mann and P. Wesson (World Scientific, Singapore)
1991.
\item S.B. Giddings and A. Strominger, Nucl. Phys. B 306 (1988) p.890.
\item T. Fukuyama and M. Morikawa, {\em Time dependence of
Coleman-Hawking Mechanism } Kyoto preprint (1988).
\end{enumerate}

\section{Figures}
Fig.1\\
An example from eq.(19) of a quantum wormhole solution.\\
 $\Psi(a)\sim exp(-a^2/2)\;_1F_1(-6;1/2;a^2)$.
 This oscillation region with
an exponential decay beyond is what we term type (I) behaviour. It
corresponds to a matter source which do not have classical wormholes
but satisfy the HP conditions.
 All Figures obtained using {\em Mathematica}.\\
Fig.2\\
As the potential is reduced the Euclidean regime reaches the origin.
We plot the modified Bessel function $\Psi(a)\sim K_{1/2}(a^2)$. This
is now the more restrictive type (II) behaviour which have classical
wormholes\\
Fig.3\\
The Wheeler-DeWitt potential $U$. The Euclidean regime has $U\geq 0$
beyond which it is Lorentzian. The tunneling boundary condition
describes the decay from the origin to $a=\Lambda^{-1/2}$.
\end{document}